\documentclass[a4paper,10pt,twocolumn]{article}

\usepackage[english]{babel}
\usepackage[utf8]{inputenc}
\usepackage[T1]{fontenc}
\usepackage[noadjust]{cite}

\usepackage[top=1.5cm, left=1.5cm, right=1.5cm, bottom=1.5cm]{geometry}

\renewenvironment{abstract}{\bf\small {\em\ Abstract---}}{}

\usepackage{amsfonts,amssymb,amsmath,amsthm}
\usepackage{subfigure}
\usepackage{graphicx}
\usepackage[footnotesize]{caption}

\usepackage{amsfonts,amssymb,amsmath,amsthm}
\usepackage{verbatim}
\usepackage{hyperref}
\usepackage[dvipsnames]{xcolor}
\usepackage{tikz}
\usetikzlibrary{spy}
\usepackage{pgfplots}
\pgfplotsset{compat=newest}
\usepackage{algpseudocode,algorithmicx,algorithm}
\usepackage{cleveref}

\usepackage{todonotes}

\usepackage{xspace}
\usepackage{bm}

\title{Efficient atom selection strategy for iterative sparse approximations}

\author{Clément Dorffer$^1$\thanks{The authors thank the DGA/MRIS, the ONR (N62909-17-1-2007) and the ANR (ANR-15-CE23-0021) for their financial support.}, Angélique Drémeau$^1$ and Cédric Herzet$^{2}$\\
  \footnotesize $^1$Lab-STICC UMR 6285, CNRS, ENSTA Bretagne, Brest, F-29200, France.\\ \footnotesize $^2$INRIA Centre Rennes-Bretagne Atlantique and Lab-STICC UMR 6285, CNRS, IMT-Atlantique, Rennes, F-35000, France.} \date{\empty} 

\newcommand{\y}{\ensuremath{\bm{y}}\xspace}
\newcommand{\R}{\ensuremath{\mathbb{R}}\xspace}
\newcommand{\dico}{\ensuremath{\mathcal{A}}\xspace}
\newcommand{\atom}{\ensuremath{\vec{a}}\xspace}
\newcommand{\subdico}{\ensuremath{\bar{\mathcal{A}}}\xspace}
\newcommand{\region}{\ensuremath{\mathcal{R}}\xspace}
\newcommand{\x}{\ensuremath{\bm{x}}\xspace}
\newcommand{\rmax}{\ensuremath{L}\xspace}
\newcommand{\rdx}{\ensuremath{l}\xspace}
\newcommand{\dicoscr}{\ensuremath{\dico_{\mathrm{removed}}}\xspace}

\newcommand{\aMl}[1]{\ensuremath{\arg\max\limits_{#1}}}

\newcommand{\normDeux}[1]{\ensuremath{\left\Vert#1\right\Vert_2}}

\newcommand{\scalprod}[2]{\ensuremath{\left\langle#1,#2\right\rangle}\xspace}
\newcommand{\abs}[1]{\ensuremath{#1}\xspace}
\renewcommand{\vec}[1]{\ensuremath{\bm{#1}}\xspace}
\newcommand{\Sphere}[2]{\ensuremath{\mathcal{B}_{#1,#2}}\xspace}
\newcommand{\Dome}[2]{\ensuremath{\mathcal{D}_{#1,#2}}\xspace}

\newcommand{\ie}{\textit{i.e.,}\xspace}

\begin{document}

\maketitle

\begin{abstract} 
We propose a low-computational strategy for the efficient implementation of the ``atom selection step'' in sparse representation algorithms. The proposed procedure is based on simple tests enabling to identify subsets of atoms which cannot be selected. Our procedure applies on both discrete or continuous dictionaries. Experiments performed on DOA and Gaussian deconvolution problems show the computational gain induced by the proposed approach. 
\end{abstract}

\section{Problem statement}

Sparsely approximating a signal vector $\y\in\R^m$ in a dictionary $\dico$ consists of finding $k\ll m$ coefficients $x_i$ and atoms $\atom_i\in\dico$ such that $\y\approx\sum_{i=1}^k\atom_i x_i$. The dictionary $\dico$ can be either discrete, \ie composed of a finite number of elements, or ``continuous'', \ie having an infinite uncountable number of atoms.

Sparse approximations have proven to be relevant in many application domains and a great number of procedures to find ``good'' sparse approximations have been proposed in the literature: convex relaxation \cite{Frank_1956,Tibshirani_1996,Daubechies_2004,Beck_2009,Rao_2015,Ekanadham_2011,Tang_2013,Xenaki_2015,Catala_2017,Boyd_2017}, greedy algorithms \cite{Mallat_1993,Pati_1993,Tropp_2006,Needell_2009,Knudson_2014}, Bayesian approaches \cite{Fevotte_2006,Schniter_2008,Soussen_2011,Dremeau_2011}, etc. 
Many popular instances of these procedures rely on the same ``atom selection'' step, \ie  the new atom added to the support at each iteration, says $\vec{a}_{\text{select}}$, verifies
\begin{equation}
    \vec{a}_{\text{select}} \in \aMl{\vec{a}\in\dico} \abs{\scalprod{\vec{r}}{\vec{a}}},
    \label{atom_sel}
\end{equation}
where $\scalprod{\vec{\cdot}}{\vec{\cdot}}$ denotes the inner product\footnote{In case of possibly negative coefficients in \x, it is common to use the absolute value of the inner product. However, one can also deal with negative coefficients using \cref{atom_sel} with a doubled size dictionary containing atoms $\vec{a}_i$ and their negatives $-\vec{a}_i$.} and $\vec{r}$ is the current ``residual'', \ie the original signal from which the contributions from the previously selected atoms have been removed. The Frank-Wolfe algorithm \cite{Frank_1956}, the matching pursuit \cite{Mallat_1993} or the orthogonal matching pursuit \cite{Pati_1993} procedures are popular instances of algorithms using \eqref{atom_sel}. It is worth noting that \eqref{atom_sel} constitutes the core of most sparse-approximation algorithms in the context of continuous dictionaries \cite{Knudson_2014,Ekanadham_2011} where some standard ``matrix-vector'' operations available in the discrete setting are no longer possible.

A brute-force evaluation\footnote{In the discrete setting, the standard approach consists of evaluating $\scalprod{\vec{r}}{\vec{a}}$ for all $\vec{a}\in\dico$, leading to a complexity scaling as $\mathcal{O}(\mathrm{card}(\dico)m)$. In the continuous setting, a classic approach consists of running gradient ascent algorithms initialized on a fine discretization of the dictionary.} of \eqref{atom_sel} may become resource consuming when the number of atoms in $\dico$ is large since it requires the exploration of the whole dictionary to find the atom the most correlated with $\vec{r}$. In this work, we propose a strategy to alleviate the complexity of \eqref{atom_sel}. Our procedure is inspired from work \cite{Herzet_2018}: it consists of performing simple tests allowing to identify group of atoms not attaining the maximum value of \eqref{atom_sel}. Interestingly, the proposed approach provides a rigorous framework to recently-proposed procedures based on some approximations of continuous dictionaries \cite{Knudson_2014,Ekanadham_2011}. In the rest of this abstract, we do not elaborate on the connections with \cite{Knudson_2014,Ekanadham_2011} (details will be provided during the conference) but rather focus on the description of the proposed methodology.



\section{Proposed strategy}\label{sec:proposed method}

Our proposed selection strategy is based on the following observations. 
If $\subdico\subseteq \dico$,  
\begin{equation}
    \max\limits_{\vec{a}\in\subdico}\abs{\scalprod{\vec{r}}{\vec{a}}}\leq\max\limits_{\vec{a}\in\dico} {\scalprod{\vec{r}}{\vec{a}}}.
    \nonumber
\end{equation}
Hence, letting $\tau \triangleq   \max\limits_{\vec{a}\in\subdico}\abs{\scalprod{\vec{r}}{\vec{a}}}$, we have $\forall \vec{a}\in \dico$:
\begin{equation}
    \abs{\scalprod{\vec{r}}{\vec{a}}}< \tau \Rightarrow
    \vec{a}\notin \aMl{\tilde{\vec{a}}\in\dico} \abs{\scalprod{\vec{r}}{\tilde{\vec{a}}}}. 
    \label{eq_major}
\end{equation}
In other words, if $\vec{a}\in\dico$ is an atom which satisfies the inequality in the left-hand side of \eqref{eq_major}, then this atom is surely not the one to be selected by \eqref{atom_sel}. 
 Elaborating on this observation, we further have:
\begin{equation}
    \max\limits_{\vec{a}\in\region}\abs{\scalprod{\vec{r}}{\vec{a}}}< \tau \Rightarrow \forall\vec{a}\in \dico\cap\region: 
    \vec{a}\notin \aMl{\tilde{\vec{a}}\in\dico} \abs{\scalprod{\vec{r}}{\tilde{\vec{a}}}}, 
    \label{eq_major2}
\end{equation}
where $\region$ is some arbitrary subset of $\R^m$. In the sequel we will refer to $\region$ as ``region''. The operational meaning of \eqref{eq_major2} is as follows: if the inequality in the left-hand side is satisfied, one is ensured that no atom in $\dico\cap\region$ will attain the maximum of $\scalprod{\vec{r}}{\vec{a}}$. The entire set $\dico\cap\region$ can thus be ignored, enabling us to reduce the number of candidate atoms to be tested in the selection \eqref{atom_sel}.


Implication \eqref{eq_major2} constitutes the basis of our complexity reduction method. More specifically, we consider \eqref{eq_major2} with some particular choices of region $\region$. These choices are motivated by the following requirements: \textit{i)} $\region$ should lead to any easy evaluation of $\max\limits_{\vec{a}\in\region}\abs{\scalprod{\vec{r}}{\vec{a}}}$; \textit{ii)} $\region$ should approximate as tightly as possible some part of $\dico$ since larger regions typically lead to inequalities more difficult to satisfy. 

In our contribution, we show that the first requirement is satisfied for some particular geometries of region $\region$. In particular, we consider ``sphere'', ``dome'' and ``slice'' geometries. Sphere and dome regions can be formally expressed as 
\begin{align}\nonumber 
\begin{array}{ll}
    \Sphere{\vec{t}}{\epsilon}\triangleq\{\vec{a}~:~\normDeux{\vec{a}-\vec{t}}\leq\epsilon\}& \mbox{(sphere)}\\
    \Dome{\vec{t}}{\epsilon}\triangleq\{\vec{a}~:~\scalprod{\vec{t}}{\vec{a}}\geq\epsilon, \normDeux{\vec{a}}= 1\}& \mbox{(dome)},
\end{array}
\end{align}
whereas the mathematical characterization of the slice regions is more involved and not detailed in this abstract. We show that for these choices of regions, $\max\limits_{\vec{a}\in\region}\abs{\scalprod{\vec{r}}{\vec{a}}}$ admits a simple analytical expression. In particular, evaluating $\max\limits_{\vec{a}\in\region}\abs{\scalprod{\vec{r}}{\vec{a}}}$ for sphere and dome regions basically requires the computation of \textit{one single} inner product $\scalprod{\vec{t}}{\vec{r}}$.

We address the second requirement by proposing a methodology to automatically adapt the size of $\region$ (via a tuning of $\epsilon$) in order to satisfy the inequality in the left-hand side of \eqref{eq_major2}. We do not detail this procedure here but mention that the evaluation of the ``optimal'' value of $\epsilon$ has a negligible complexity. 

We thus propose the following strategy (summarized in Algorithm~\ref{algo}) to speed up the computation of \eqref{atom_sel}. We select a set $\subdico\subset\dico$ and $\rmax$ regions $\{\region_\rdx\}_{l=1}^\rmax$, 
and apply test \eqref{eq_major2} for each region. Each test allows to identify a set\footnote{Which may be empty in some cases.} of atoms which do not attain the maximum of  $\scalprod{\vec{r}}{\vec{a}}$. We evaluate \eqref{atom_sel} by working on a reduced dictionary:
\begin{equation}
        \vec{a}_{\text{select}} \in \aMl{\vec{a}\in\dico\backslash \dicoscr} \abs{\scalprod{\vec{r}}{\vec{a}}},
    \label{atom_selred}
\end{equation}
where $\dicoscr$ denotes the set of atoms which have been removed by the tests \eqref{eq_major2}.   

For discrete dictionaries, the computational complexity of the proposed method (for sphere and dome regions) evolves as
\begin{equation}
        \mathcal{O}(\underbrace{\mathrm{card}(\subdico) m}_{\mbox{evaluation of $\tau$}} 
        + \underbrace{\rmax m}_{\mbox{evaluation of \eqref{eq_major2}}}
        + \underbrace{\mathrm{card}(\dico\backslash \dicoscr)m}_{\mbox{evaluation of \eqref{atom_selred}}}).\nonumber 
\end{equation}
This has to be compared to the complexity required by a brute evaluation of \eqref{atom_sel}, \ie $\mathcal{O}(\mathrm{card}(\dico)m)$.
In the next section, we propose to compare the efficiency and the computational gain allowed by the proposed methodology. 

\begin{algorithm}[t]
\caption{Efficient selection strategy}
\label{algo}
\begin{algorithmic}
\State \textbf{Input:} residual \vec{r}, subset of atoms \subdico, set of regions $\{\region_\rdx\}_{\rdx=1}^\rmax$
\State Init: : $\dicoscr = \emptyset$
\State  Evaluate $\tau = \max\limits_{\vec{a}\in\subdico}\abs{\scalprod{\vec{r}}{\vec{a}}}$
\ForAll {$1\leq \rdx\leq \rmax$}
                \If {$\max\limits_{\vec{a}\in\region_\rdx}\abs{\scalprod{\vec{r}}{\vec{a}}}< \tau$}
                \State $\dicoscr = \dicoscr \cup (\dico\cap\region_\rdx)$
                \EndIf
\EndFor
\State Find $\vec{a}_{\text{select}} \in \aMl{\vec{a}\in\dico\backslash \dicoscr} \abs{\scalprod{\vec{r}}{\vec{a}}}$
\end{algorithmic}
\end{algorithm}

\section{Experiments}

\begin{figure}[t]
\centering
\subfigure{
\resizebox{.8\linewidth}{!}{\begin{tikzpicture}
\begin{semilogyaxis}[ legend cell align={left},legend style={at={(0.4,.8)}, anchor=south} ,ybar , enlargelimits=false, width=\linewidth,ylabel={\# of scal. prod.}, xlabel={Iterations},xmin=.5,xmax=5.5,xtick={1,2,3,4,5},ymin=10,ymax=50000,ytick={1.e2,1.e3,1.e4},line width=.25pt,bar width=2pt,title =(a)]
\addplot[very thin,fill,color=blue] table[x=iter,y=SP_OMP] {SP_DoA.dat};
\addplot[very thin,fill,color=red] table[x=iter,y=SP_SOMP] {SP_DoA.dat};
\legend{classical exhaustive selection,proposed strategy}
\end{semilogyaxis}
\end{tikzpicture}}
\label{fig_DoA}}
\subfigure{
\resizebox{.8\linewidth}{!}{
\begin{tikzpicture}[spy using outlines={circle,black,magnification=7.5,size=3cm, connect spies}]
\begin{axis}[xlabel={Region centers},ylabel={"Screened" interval},title=(b)]
\addplot [color=blue, only marks,mark = +, mark size=1pt]
plot [error bars/.cd, y dir = both, y explicit]
table[x =a, y =center, y error =radius]{interv_deconv_v2.dat};
\end{axis}
\spy on (5.65,4.55) in node[fill=white] at (5.7,1.75);
\end{tikzpicture}}\label{fig_deconv}}
\caption{(a) : ``discrete'' DOA estimation problem - number of scalar products evaluated for the atom selection. (b) : ``continuous'' Gaussian deconvolution problem - interval of atoms which can be ignored in problem \eqref{atom_sel}. }
\label{fig}
\end{figure}
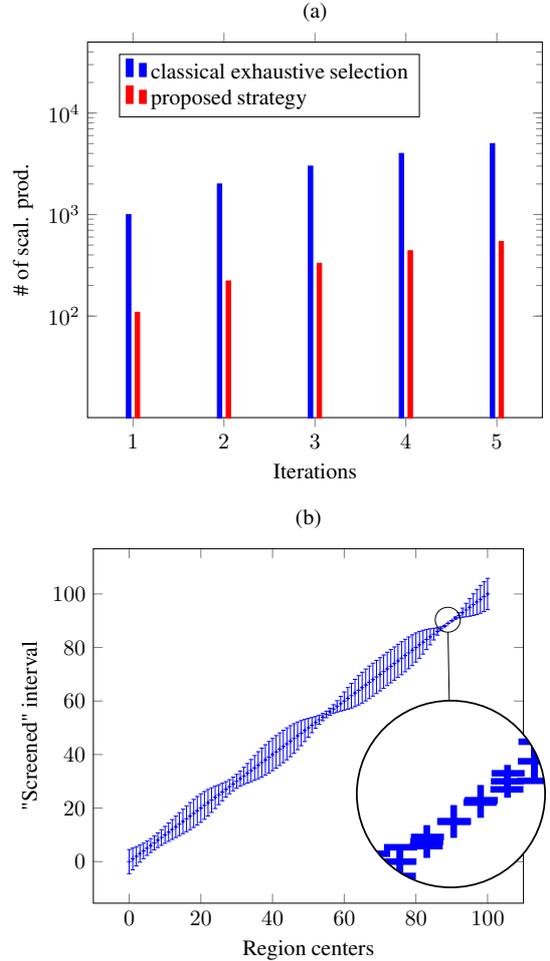

We propose to challenge the proposed selection procedure and the classical exhaustive search on two different problems: a direction-of-arrival (DOA) estimation problem using a discrete dictionary and a Gaussian deconvolution problem with a continuous dictionary. Within the DOA estimation framework, we examine the number of scalar products required to achieve the selection step \eqref{atom_sel}, with and without the proposed method. This provides a quantitative assessment of the computational gain induced by our method. The Gaussian deconvolution problem acts as a proof of concept, illustrating the interest of our method in continuous dictionaries. 

 In the DOA estimation problem, we consider a dictionary composed of $n=1000$ normalized steering vectors of size $m=100$, each corresponding to a different angle of arrival in $[-\pi/2,\pi/2]$. In the Gaussian deconvolution problem, $\dico = \{a(\mu) : \mu \in [0,100]\}$ where $a(\mu)$ is a Gaussian function with mean $\mu$ and variance $\sigma^2=10$. In both simulations, the observed signal $\vec{y}$ is constructed as a random linear combination of $5$ atoms of \dico. 
Coefficients associated to the atoms are realizations of a uniform distribution on $[-1,1].$

In Fig.~\ref{fig_DoA}, we considered the DOA estimation problem. We applied 5 iterations of orthogonal matching pursuit on $\vec{y}$ (which requires to solve \eqref{atom_sel} at each iteration). 
Each column of the figure represents the (cumulated) number of inner products which have been computed with an exhaustive search of $\vec{a}_{\text{select}}$ in the whole dictionary (blue), and with the proposed method described in Algorithm~\ref{algo} (red).  We set $\rmax = 100$ and define the centers of the (dome) regions $\{\vec{t}_\rdx\}_{\rdx=1}^\rmax$ by a regular subsampling of the dictionary. We let $\subdico=\{\vec{t}_\rdx\}_{\rdx=1}^\rmax$. We see that the proposed method allows for a gain of complexity of one order of magnitude with respect to a brute-force approach. 


In Fig.~\ref{fig_deconv} we consider the Gaussian deconvolution problem and show the set of removed atoms, $\dico\cap\region_\rdx$, for each (sphere) region $\{\region_\rdx\}_{\rdx=1}^{\rmax = 100}$. We choose the centers of the regions $\{\vec{t}_\rdx\}_{\rdx=1}^\rmax$ by a regular subsampling of $\mu\in\left[0,100\right]$. The size of each region (value of $\epsilon$) is automatically tuned to verify (if possible) the inequality in the left-hand side \eqref{eq_major2} as discussed in Section~\ref{sec:proposed method}. We set $\subdico = \{\vec{t}_\rdx\}_{\rdx=1}^{\rmax=100}$. The figure shows the interval of value of $\mu$ which can be removed from the dictionary for each test region.  
We see in Fig.~\ref{fig_deconv} that by the end of the proposed elimination procedure, the search space is reduced to a small interval, around $\mu\approx90$.


\bibliographystyle{unsrt}
\bibliography{biblio.bib}

\end{document}